# Dielectric spectroscopy in benzophenone: The $\beta$ relaxation and its relation to the mode-coupling Cole-Cole peak

L.C. Pardo[1,2], P. Lunkenheimer[1,*], and A. Loidl[1]

[1]*Experimental Physics V, Center for Electronic Correlations and Magnetism, University of Augsburg, 86135 Augsburg, Germany*
[2]*Departament de Física i Enginyeria Nuclear, ETSEIB, Universitat Politècnica de Catalunya, Diagonal 647, 08028 Barcelona, Catalonia, Spain*

We report a thorough characterization of the glassy dynamics of benzophenone by broadband dielectric spectroscopy. We detect a well pronounced $\beta$ relaxation peak developing into an excess wing with increasing temperature. A previous analysis of results from Optical-Kerr-effect measurements on this material within the mode coupling theory revealed a high-frequency Cole-Cole peak. We address the question if this phenomenon also may explain the Johari-Goldstein $\beta$ relaxation, a so far unexplained spectral feature inherent to glass-forming matter, mainly observed in dielectric spectra. Our results demonstrate that according to the present status of theory, both spectral features seem not to be directly related.

Aside of the $\alpha$ relaxation, characterizing the slowing down dynamics during the glass transition, glassy matter reveals a rich variety of further, so far only poorly understood dynamic processes [1,2,3]. Among the most prominent ones is the so-called Johari-Goldstein (JG) $\beta$ relaxation [4]. In spectra of the imaginary part of the susceptibility $\chi''$ (the "loss") it shows up as second peak at a frequency beyond the $\alpha$ relaxation. It is inherent to the glassy state of matter but its microscopic origin still is controversially discussed. Even if superimposed by a dominating $\alpha$ relaxation, it still may show up in the loss as a so-called excess wing [5]. This long known feature [2,3,6,7] only recently was discovered to be due to a second relaxation peak [8].

The most prominent theory of the glass transition is the mode coupling theory (MCT) [9]. It predicts a so-called fast $\beta$ relaxation, showing up in the GHz - THz range, which was experimentally verified in numerous investigations [10]. It is envisioned by the rattling motion of a particle in the cage formed by the surrounding particles. It should be noted that this fast $\beta$ process usually is presumed not to be identical with the JG $\beta$ relaxation, sometimes also termed "slow $\beta$ process". In the basic versions of MCT, a combination of two asymptotic power laws is predicted to describe the data in the regime of the fast process, with both exponents determined by a system parameter $\lambda$. They form a shallow minimum in the frequency-dependent loss, followed by the so-called microscopic peak located in the THz range [Fig. 1(a)]. Some consensus seems to develop in the glass community that MCT is the correct description for high temperatures, $T > T_c$. The critical temperature $T_c$ can be regarded as idealized glass-transition temperature often roughly located at 1.2 $T_g$ with $T_g$ the experimental glass temperature. The original MCT only revealed three characteristic spectral features, the $\alpha$ relaxation, the minimum, and the microscopic peak. The high-frequency flank of the $\alpha$ relaxation peak directly crosses over into the low-frequency wing of the minimum, the so-called von-Schweidler law $\nu^{-b}$ [Fig. 1(a)]. At high temperatures this is in accord with experimental observations. However, excess wing or JG $\beta$ peak developing at lower temperatures seemed not to be covered by MCT.

Recently it was shown that the asymptotic power laws of basic MCT could not describe optical Kerr effect (OKE) results on benzophenone (BZP) [11,12], even at $T > T_c$. However, by accounting for rotation-translational coupling in a schematic model, Götze and Sperl demonstrated that these data indeed are consistent with MCT [13,14]. Based on earlier theoretical work [15] they pointed out that the fast $\beta$ relaxation of MCT can lead to a peak that is approximately described by the empirical Cole-Cole (CC) law [16], with flanks $\nu^g$ and $\nu^a$. If the CC peak frequency is larger than that of the microscopic excitations, the conventional case of a loss minimum arises [Fig. 1(a)]. However, the CC peak also can be located at much smaller frequency and instead lead to a second more shallow power law $\nu^{-a}$ at the high-frequency flank of the $\alpha$ peak [Fig. 1(b)]. The OKE results in BZP indeed could be fitted assuming the latter case. Very recently depolarized light scattering (DLS) experiments on BZP were reported [17], in good accord with the OKE results. The loss spectra from OKE [12] and DLS, presented there, indeed revealed such a power law.

The CC function also commonly is used for the description of the JG $\beta$ relaxation. In addition, it was stated [14,17,18] that the OKE and DLS spectra in BZP resemble typical dielectric spectra with excess wing, found in various glass formers. Thus it is suggestive to assume that the CC peak of MCT, located at about 10 GHz in BZP [14], is related to the JG process commonly detected by dielectric spectroscopy. This would imply MCT covering all processes of glassy dynamics, including the slow $\beta$ process. However, usually JG $\beta$ relaxation and excess wing are reported at lower temperatures and frequencies where they become most pronounced [2,6,7]. The excess wing even can develop into a slow $\beta$ peak when approaching $T_g$ [8]. Therefore it is suggestive to investigate BZP at lower frequencies and temperatures using dielectric spectroscopy. It is known since long that BZP can be supercooled [19] and $T_g = 212$ K was reported [20]. However, so far there is no thorough dielectric investigation of this glass former.

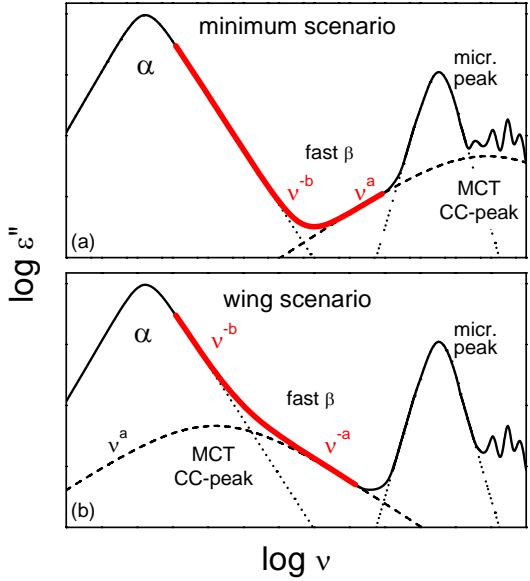

FIG. 1 (Color online). Schematic loss spectra for two scenarios of MCT. Depending on the frequency of the CC peak, the conventional case of a fast $\beta$ minimum (a) or a wing at the right flank of the $\alpha$ peak (b) can arise [13,14].

Benzophenone (diphenylketone, $H_5C_6-CO-C_6H_5$) was purchased from Merck with a minimum purity of 99% and used without further purification. Additional measurements in distilled material revealed identical results. For details on the measurement techniques used up to 3 GHz the reader is referred to [21]. In the region 300 MHz $\leq \nu \leq$ 30 GHz an Agilent E8363 network analyzer was used to measure the reflection of a coaxial line whose end was immersed into the sample material [22]. For cooling/heating a nitrogen gas-heating system and an isothermal heat bath were used. In the region of 230 K - 255 K, BZP exhibits an enhanced crystallization tendency. Spectra at $T \leq 235$ K were collected under heating after passing this region with rapid cooling rates of up to 10 K/min. Spectra between 240 K and 250 K were measured during quickly cooling the sample with rates up to 3 K/min. Thus in this region the thermal equilibration of the sample may not have been perfect and a temperature error up to 1 K has to be assumed.

Figure 2 shows the dielectric loss spectra of BZP for various temperatures. Well-pronounced $\alpha$ relaxation peaks show up, their continuous temperature-dependent shift reflecting the glassy freezing of molecular dynamics. The empirical Cole-Davidson (CD) function [23] provides reasonable fits of the spectra (solid lines). At $T \leq 230$ K, a Cole-Cole (CC) function, eq. (1), was added to the fitting function to account for the excess wing and $\beta$ relaxation as detailed below. Fig. 3 shows $\tau_\alpha(T)$, significantly extending the so far investigated range [11,12,17] to lower temperatures. It exhibits distinct non-Arrhenius behavior. We used the Vogel-Fulcher-Tammann (VFT) function, $\tau_\alpha = \tau_0 \exp[DT_{VF}/(T-T_{VF})]$ with $T_{VF}$ the Vogel-Fulcher temperature and $D$ the strength parameter [24], to parameterize the data (solid line in Fig. 3). Within the strong-fragile classification scheme by Angell [24], the small value of $D = 3.8$ characterizes BZP as fragile glass former. As often found [25], small but significant deviations of fit and experimental data show up. The deviations at sub-$T_g$ temperatures are due to the loss of thermodynamic equilibrium. Fig. 3 also includes relaxation times from the OKE experiments as reported in [12] and as deduced from the peak positions $\nu_p$ of the DLS susceptibility provided in [17]. There is a reasonable agreement with small deviations at high temperatures, which may be understood considering the different tensorial properties of the methods [26,27]. In [12] good agreement of $\tau_\alpha(T)$ with the critical law predicted by basic MCT, namely $\tau_\alpha \sim (T-T_c)^{-\gamma}$ with $\gamma = 1/(2a) + 1/(2b)$ [9], was reported. The dashed line shown in Fig. 2 is a fit of the dielectric data at $T > 255$ K using the same $T_c$ and $\gamma$ as in [12]. The revealed deviations at $T < T_c$, also found in other glass formers [2], can be understood within extended versions of MCT [9]. The inset of Fig. 2 (cf. Fig. 4 in [12]), demonstrates that critical behavior is well obeyed at high temperatures, $T > T_c$. Also the width parameter of the α relaxation, tending to saturate at a high temperature value below unity, and the relaxation strength, showing a cusp-like anomaly at $T_c$, in most respects are consistent with MCT [28]. Thus overall the MCT predictions concerning the α relaxation are well fulfilled by the dielectric spectroscopy results on BZP. A more detailed discussion of the α relaxation behavior of BZP will be given in [28].

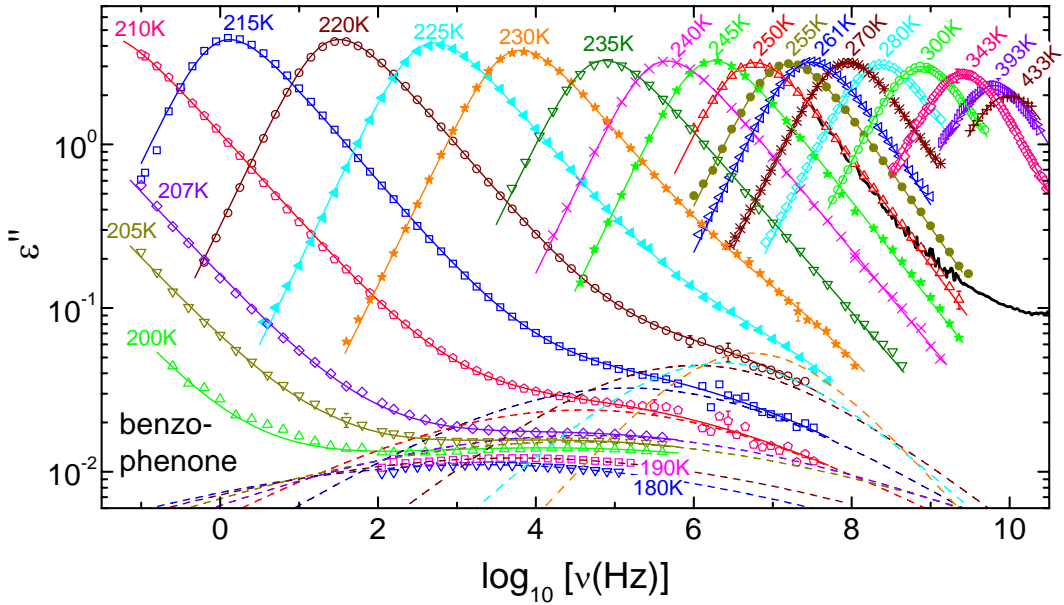

FIG. 2 (Color online). Frequency-dependent dielectric loss of BZP for various temperatures. The solid lines are fits with the sum of a CD and a CC function (at $T \leq 190$ K the CD amplitude and at $T \geq 235$ K the CC amplitude were set to zero). The dashed lines show the CC part of the fits accounting for the observed $\beta$ relaxation. The thick solid line superimposed to the 250 K spectrum is the OKE susceptibility at 251 K [12,17]. It was vertically scaled to match the dielectric data at 250 K.

It is clearly revealed by Fig. 2 that BZP has a β relaxation in the Hz - MHz range. It shows typical features of a canonical JG β relaxation: It can be well fitted by the CC function (dashed lines), its peak amplitude increases and the peak width decreases with increasing temperature. At high temperatures the β peak first develops into an excess wing and finally becomes completely submerged under the dominating α peak. This happens already at 235 K, significantly below the lowest temperature investigated in the OKE and DLS experiments (251 K) where the most pronounced excess-wing like feature showed up [11,12,17]. Despite the restricted frequency range of the dielectric spectrum, direct comparison with the OKE result at this temperature (thick solid line in Fig. 2) reveals a significant deviation: The excess intensity at $\nu > 1$ GHz in the OKE spectrum, which is assumed to arise from the MCT CC peak [14], is not seen in $\varepsilon''(\nu)$. It is a well established fact that usually the minimum caused by the fast β process of MCT has much smaller relative weight in dielectric spectra than in those obtained by scattering methods [2,29]. This can be understood, at least qualitatively, taking into account different coupling to density fluctuations and different tensorial properties of the experimental probes [26,29,30]. The recently established CC peak of MCT [13,14] is due to just the same fast β dynamics as the minimum and thus also here a smaller amplitude in dielectric spectra can be expected. This could easily explain the differences in the 250/251 K spectra and the fact that the β relaxation in $\varepsilon''(\nu)$ is already submerged at 235 K. However, it should be noted that in a recent work comparing the slow β relaxation in light scattering and dielectric spectra [31] a much smaller amplitude of this feature in light scattering was found.

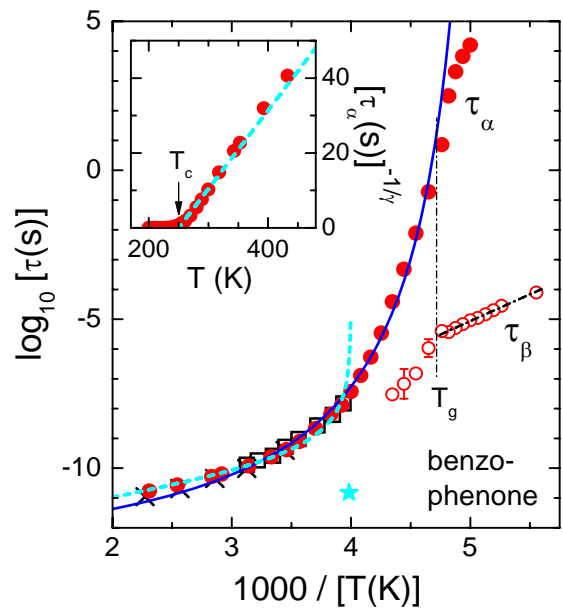

Fig. 3 (Color online). Relaxation times determined from the fits shown in Fig. 2 (circles). In addition, data from OKE (squares) [12] and DLS (x's) [17] experiments are shown. The solid line is a fit of the dielectric data with the VFT function ($\tau_0 = 4.2 \times 10^{-13}$ s, $T_{VF} = 189$ K, $D = 3.8$). The dashed line is a fit at $T \geq 255$ K with MCT (see text), using $T_c = 250$ K and $\gamma = 1.92$ as reported in [12]. The dash-dotted line is a fit of the β relaxation time at $T < T_g$ assuming an Arrhenius law ($\tau_0 = 8.8 \times 10^{-15}$ s, $E = 0.36$ eV). The star shows the CC relaxation time as determined by Sperl [14] from a MCT analysis of the OKE data. The inset shows the dielectric data and MCT fit, using a representation that should linearize according to MCT ($\gamma = 1.92$ [12]).

The β relaxation times $\tau_\beta(T)$ obtained from the fits are shown in Fig. 3 (open circles). Arrhenius behavior of $\tau_\beta(T)$ as revealed at $T < T_g$ in BZP, is often considered as typical property of JG β relaxations. However, one should be aware that this notion is mainly based on measurements below $T_g$, where even the α relaxation time follows an Arrhenius law (see, e.g., uppermost three points of $\tau_\alpha$ in Fig. 2). Above $T_g$, $\tau_\beta(T)$ crosses over to much stronger non-Arrhenius temperature dependence as also found in other glass formers [3,8,32]. It should be noted that especially at elevated temperatures the determination of $\tau_\beta$ has high uncertainty due to the strong overlap with the α process (see error bars in Fig. 2). Also one may doubt if a simple *additive* superposition of α and β relaxation is valid in this region [33]. However, an alternative, non additive superposition [33] led to results consistent with those shown in Fig. 2.

The single star shown in Fig. 3 is the relaxation time determined from the peak frequency of the CC peak (10.7 GHz),

which was obtained from the MCT analysis of the 251 K OKE results [14]. It seems very unlikely that the $\tau_\beta(T)$ curve from the dielectric data should proceed to this value. In this context it is interesting, that according to eqs. (16a) and (16b) of ref. [14], the CC peak frequency should be smaller for reduced CC-peak amplitude. As discussed above, this amplitude should vary for different experimental techniques. Thus the CC peak frequency and the corresponding relaxation time may well depend on the method used to detect it. Judging from earlier investigations, the amplitude in the fast $\beta$ process regime can be up to one order of magnitude smaller for dielectric spectroscopy compared to DLS [2,26,29]. Therefore for dielectric spectroscopy on BZP the CC relaxation time of MCT may be expected at a one decade higher value than the star shown in Fig. 3. However, even then an extrapolation of the experimentally detected points (open circles) to such a value seems difficult. In fact, according to the current status of the theory, the CC peak should not or only weakly shift along the frequency axis with temperature and thus the associated relaxation time should be constant [14]. In [14] the temperature variation of the OKE data could be taken into account by the temperature dependence of the von Schweidler law alone, without any shift of the CC peak. This is not possible for the much stronger variation of the $\beta$ peak revealed by the broader temperature range of the dielectric experiment.

The width of the observed symmetric $\beta$ relaxation peaks varies markedly with temperature (Fig. 2) and the frequency exponent of their flanks increases from 0.15 to 0.55. In the MCT framework outlined in [13,14], this exponent is identified with the so-called critical exponent $a$, which should be temperature independent [9]. However, the experimental data reveal strong temperature dependence, as commonly observed for slow $\beta$ relaxations. In addition, at $T > 220$ K, the spectra cannot be fitted with $a < 0.4$, which is the allowed maximum value for this quantity within MCT [9].

In conclusion, our broadband measurements characterize BZP as a typical fragile glass former that has a well pronounced slow JG $\beta$ relaxation at low temperatures. The $\alpha$ relaxation in most respects is in good accord with MCT predictions [28]. Concerning the excess-wing like feature seen by OKE/DLS and analyzed by the MCT CC-peak, it is clear from our findings that it cannot be explained by a simple continuation of the detected $\beta$ relaxation to higher temperatures and frequencies. If there is a connection of both features, the situation must be more complicated. The difference of the spectra at 250/251 K unequivocally demonstrates that different coupling of the different methods to the fast MCT process has to be taken into account. Clearly the detected $\beta$ relaxation in BZP cannot be described with a temperature-independent CC-peak. Thus, overall we can state that according to the current status of the theory, the CC peak of MCT used to describe the high-frequency OKE data cannot be due to the canonical $\beta$ relaxation as detected by dielectric spectroscopy in this material. Only if future theoretical developments could capture the strong temperature dependence of the CC peak-frequency and width, the final goal of explaining all dynamic processes of glass formers, including slow $\beta$ relaxation and excess wing, may be achieved.

We thank M. Sperl for stimulating discussions and helpful comments and E.A. Rössler for providing the files of the DLS and OKE data shown in Fig. 3. One of the authors (L.C.P.) would like to acknowledge the financial support of the Humboldt Foundation, Ministerio de Educación y Ciencia (grant FIS2005-00975 and post doctoral fellowship MEC-EX2005), and of Generalitat de Catalunya (grant SGR2005-00535).